# Radial thermal expansion of single-walled carbon nanotube bundles at low temperatures.

A.V. Dolbin[1], V.B. Esel'son[1], V.G. Gavrilko[1], V.G. Manzhelii[1], N.A.Vinnikov[1], S.N. Popov[1], B. Sundqvist[2].

[1] B. Verkin Institute for Low Temperature Physics & Engineering NASU, Kharkov 61103, Ukraine

[2] Department of Physics, Umea University, SE - 901 87 Umea, Sweden

Electronic address: dolbin@ilt.kharkov.ua

**Abstract**.

The linear coefficient of the radial thermal expansion has been measured on a system of SWNT bundles in an interval of 2.2 – 120K. The measurement was performed using a dilatometer with a sensitivity of $2 \cdot 10^{-9}$ cm. The cylindrical sample 7 mm high and 10 mm in diameter was obtained by compressing powder. The resulting bundles of the nanotubes were oriented perpendicular to the sample axis. The starting powder contained over 90% of SWNTs with the outer diameter 1.1 nm, the length varying within 5-30 μm.

Since the discovery of carbon nanotubes in 1991 [1], this novel class of physical objects has been attracting immense experimental and theoretical interest. However, a wide variety of types of carbon nanotubes and the problems encountered in preparing pure samples of these types of nanotubes makes it extremely difficult to reveal regularities in the behavior of nanotubes (e.g., see [2] and the References in it). The thermal expansion of nanotubes remains among the least studied properties of carbon nanotubes. For example, the thermal expansion of single-walled nanotubes and their bundles has never been investigated experimentally below room temperature. Meanwhile, investigations at low temperatures can provide the most valuable information about the dynamics of nanotubes. The thermal expansion coefficients (TECs) predicted theoretically [3 – 8] for single-walled nanotubes (SWNT) differ in order of magnitude and sign.

In this study the radial thermal expansion of bundles of closed end SWNTs (c-SWNT) was measured in an interval of 2.2 – 120 K.

The sample for measuring thermal expansion was prepared using the effect of aligning the SWNT axes by pressure of 1 GPa (see [9]). Under this pressure the nanotubes within a 0.4 mm thick layer are aligned in the plane normal to the pressure vector, the average deviation from the plane of alignment being ~ 4° [9].

The starting material for the sample was a carbon nanotube powder from "Cheap Tubes", USA. According to the Raman analysis performed by the producer, the powder contained over 90% of SWNTs with the average outer diameter of the nanotubes 1.1 nm.

The main characteristics of the powder are shown in the Table. The chirality distribution of the nanotubes in the powder in unknown.

Table. The characteristics of the SWNT powder

| Diameter | 1—2 nm |
|---|---|
| Length | 5—30 μm |
| SWNT fraction | >90 wt% |
| Amorphous carbon fraction | <1.5 wt% |
| Co catalyst fraction | 2,9 wt% |
| Specific surface | >407 m$^2$/g |
| Electrical conduction | >10$^2$ S/cm |

A cylindrical sample was compressed of pressure (1.1 GPa) – oriented SWNT plates, each being up to 0.4 mm thick. The sample was 7.2 mm high and 10 mm in diameter. Its density was 1.2 g/cm$^3$. It was prepared in a special dismountable cylindrical matrix for compressing nanotube powder under the effective pressure 0.5 – 2 GPa. The matrix was made in the shape of a cone and had a ring with a cylindrical channel inside. The conic part was inserted into an outer hardened steel cylinder, which produced resistance to the internal pressure.

The inner ring consisted of four sections. After compression was completed, the interior part of the matrix was removed carefully (to preclude destruction of the sample) from the inner ring. Then the four sections of the inner ring could be detached with the minimal harm for the sample. The piston was made of stainless steel. The sample prepared by this technique is expected to have pronounced anisotropy of the properties in the directions perpendicular and parallel to the sample axes. In such a sample the axes of the nanotube bundles are disordered in the direction perpendicular to the applied pressure vector. However, in the parallel direction to sample axis the thermal expansion of the sample is determined only by the radial component of the SWNT bundles oriented in the plane.

The radial thermal expansion of the sample was investigated using a low temperature capacitance dilatometer. The dilatometer design and the measurement technique are detailed elsewhere [10]. The linear thermal expansion coefficient (LTEC) was measured in the direction coinciding with that of the pressure applied to compress the sample, i.e. radially to the bundles of nanotubes. To remove the gas impurities, the sample was subject to dynamic evacuation under 1·10$^{-3}$ mmHg for three days at room temperature. Immediately ahead of dilatometric investigation the measuring cell with the sample was cooled slowly (for 8 hours) down to liquid helium temperature (4.2K) and kept at this temperature for about 4 hours. The cooling and the subsequent investigation were made in vacuum up to $1 \cdot 10^{-6}$ mm Hg. The temperature dependence LTEC measured at T=2.5 – 120 K is shown in the figure. The $\alpha_r(T)$ values are averaged over several measurement series.

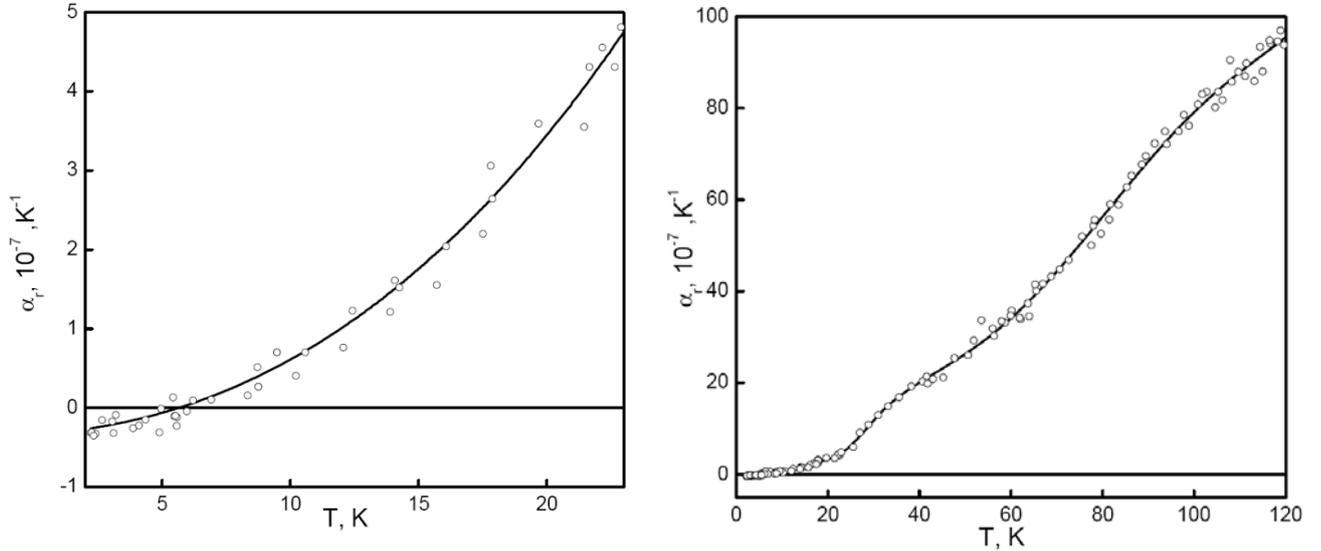

Fig. The LTEC of a sample of pressure – oriented nanotubes in the direction radial to the axes of nanotube bundles: a) T=2.5 – 120K, b) T=2.2 – 25K.

It is seen that the radial LTEC $\alpha_r$ is positive above 5.5K and negative at lower temperatures. The $\alpha_r$ – magnitude actually consists (the impurity effect is disregarded) of two components $\alpha_d$ and $\alpha_g$ accounting for the changes in the nanotube diameters and the intertube gap, respectively. These components were sole measured by the x-ray diffraction method in the interval 300 – 950K in [11] : at T=300K $\alpha_r = (0.75\pm0.25) \cdot 10^{-5}$ K$^{-1}$, $\alpha_d = (-0.15\pm0.20) \cdot 10^{-5}$ K$^{-1}$ and $\alpha_g = (4.2\pm1.4) \cdot 10^{-5}$ K$^{-1}$. In [12] applying also x-ray diffraction method $\alpha_r$ appeared to be negative in the whole interval of the experiment (290 – 1600K).

No other information about experimental investigation of the thermal expansion of SWNT bundles has come to our notice.

Neither can we compare our results with theory. Firstly, the available theoretical studies consider radial and axial thermal expansions of individual nanotubes and offer only general speculations on the effect of the intertube interaction in bundles upon thermal expansion (e.g., see [8]). Secondly, the theoretical conclusions about the magnitude, sign and temperature dependence of the thermal expansion coefficients of nanotubes, the effect of nanotube chirality and diameter upon thermal expansion and about the correlation between the radial and axial components of the thermal expansion of nanotubes are rather controversial. For example, the thermal expansion is negative in a wide temperature interval (0 – 800K) in [4], changes from a negative value at low temperatures to a positive one at moderate and high temperatures in [8] and remains positive at all the temperatures in [6].

Experimental data on the thermal expansion of nanotubes, especially in the low temperature region where the temperature dependence of thermal expansion is quite strong, can greatly improve the current theoretical models of nanotube dynamics.

The authors are indebted to the Science and Technology Center of Ukraine (STCU) for the financial support of this study (Project 4359 and 4266).